\documentclass[twocolumn,superscriptaddress,showpacs,showkeys,preprintnumbers,amsmath,amssymb,aps]{revtex4-1}
\usepackage{graphicx} % Include figure files
\usepackage{float}
\usepackage{color}
\usepackage{nicefrac}
\usepackage{collref}

\begin{document}

%\title{Topological phase transition in 3D}
\title{Topological phase transition induced by random substitution}

\author{Stanislav~Chadov}\email{stanislav.chadov@cpfs.mpg.de}
\affiliation{Max-Planck-Institut f\"ur Chemische Physik fester Stoffe, 01187  Dresden, Germany}
\author{Janos~Kiss}
\affiliation{Max-Planck-Institut f\"ur Chemische Physik fester Stoffe, 01187  Dresden, Germany}
\author{Claudia~Felser}
\affiliation{Max-Planck-Institut f\"ur Chemische Physik fester Stoffe, 01187  Dresden, Germany}
\affiliation{Institut f\"{u}r Anorganische und Analytische Chemie,  Johannes-Gutenberg-Universtit\"{a}t,  55099 Mainz Germany}
\author{Kristina~Chadova}
\affiliation{Dept. Chemie, Ludwig-Maximilians-Universit\"at, 81377
  M\"unchen, Germany}
\author{Diemo K\"odderitzsch}
\affiliation{Dept. Chemie, Ludwig-Maximilians-Universit\"at, 81377
  M\"unchen, Germany}
\author{Jan Min\'ar}
\affiliation{Dept. Chemie, Ludwig-Maximilians-Universit\"at, 81377 M\"unchen, Germany}
\author{Hubert Ebert}
\affiliation{Dept. Chemie, Ludwig-Maximilians-Universit\"at, 81377 M\"unchen, Germany}



\begin{abstract}
The transition from topologically nontrivial to a trivial state is studied
by first-principles calculations on bulk zinc-blende type
(Hg$_{1-x}$Zn$_x$)(Te$_{1-x}$S$_x$) disordered alloy series.
The random chemical disorder was treated by means of the Coherent Potential Approximation.
We found that although the phase transition occurs at the strongest disorder 
regime (${x\approx 0.5}$), it is still manifested by well-defined Bloch states
forming a clear Dirac cone at the Fermi energy of the bulk disordered material.
The computed residual resistivity tensor confirm the topologically-nontrivial
state of the HgTe-rich (${x<0.5}$), and the trivial state of the ZnS-rich 
alloy series (${x>0.5}$) by exhibiting the quantized behavior of the 
off-diagonal spin-projected component, independently on the concentration $x$.
\end{abstract}

\maketitle 

%------------------------------------------------------------------------------
\section{Introduction}
\label{sec:intro}
%------------------------------------------------------------------------------
The most attractive property of topologically-nontrivial insulating
materials is the presence of the electronic chiral states at their surface, 
forming the so-called Dirac cone, which gives rise to the quantized
non-dissipative and fully spin-polarized surface current~\cite{PHK+12}.
This surface property is exploited in band structure calculations, which 
intend to prove the trivial or nontrivial topology of a given system~\cite{BHZ06}.
Since the topology is directly related to the bulk band structure,
 those approaches which can predict the topological class of a given
material based on purely bulk information are more practical as being
computationally less  demanding compared to realistic surface calculations.
Thus, in the search for novel materials such theoretical approaches where
the topological phase transition can be straightforwardly investigated in
the bulk are very useful tools, that are able to provide supporting data to
motivate experimental investigations on unexplored new compounds.\\
\indent
Indeed, from the point of view of the band structure, the manifestation of the
trivial-nontrivial phase transition is always marked by the existence of a Dirac
point, independently on the nature of the order parameter responsible for the
assignment of a given material to a certain topological class.
In contrast to expensive and tedious experimental methods, in theoretical
studies these parameters can be easily tuned.
For example, the trivial-nontrivial phase transition can be simply induced by
changing the amplitude of the spin-orbit coupling, or the crystal-field
splitting (e.\,g.\@ through the change of the lattice constant)~\cite{CQK+10}.
In case of well-defined Bloch states one can also make use of the parity
analysis of the eigenstates~\cite{FK07}.
However, the most problematic cases are those, where the local translational
symmetry is broken due to random fluctuations of certain degrees of freedom, 
which are  inevitably always present in real materials.
These fluctuations can be caused by random chemical disorder, various dynamical
fluctuations as e.\,g.\@ phonons or magnons, or strong local electron
correlations which are specific for systems with strong spin-orbit coupling.
All these mechanisms lead to a localization of the electronic states which
can no longer be expressed in the form of Bloch waves.
This makes the straightforward analysis of the parity of the eigenstates
inapplicable.
Also, in the ordered class of materials so far only a few stable non-trivial 
compounds have been found. Therefore, the search for topological insulators among 
disordered systems will gain an increased importance.
This will be especially true for materials with application relevance, since
disordered systems are easier to be mass produced, and despite structural
disorder their topological edge states are robustly protected against any
time-reversal symmetric perturbation.
Hence, by extending our field of interest towards disordered systems many new 
topologically-nontrivial materials can be found.\\
\indent
The practical tools for the first-principles studies on the disordered
systems are typically provided by the effective mean-field theories,
such as e.\,g.\@ the Dynamical Mean-Field Theory (DMFT)~\cite{KSH+06} or
via the Coherent Potential Approximation (CPA)~\cite{Sov67,But85}.
The former takes into account the local dynamical electronic correlations,
and the latter efficiently describes the random chemical disorder.
Each implementation of the mean-field theory incorporates the effect of a 
certain type of random fluctuations in the form of an energy-dependent 
complex-valued local potential (the so-called self-energy) added to the 
real-valued lattice-symmetric Kohn-Sham potential.
This leads to an energy-dependent shift and to a broadening of the
electronic states, seen in the band structures  calculated with
DMFT or CPA.\\
\indent
In the following we would like to demonstrate the power of the CPA as a
practical adiabatic technique, which allows to determine the topological class
of a given compound by its gradual transformation into another system, for which
the topological class is already well established.
In contrast to complicated surface calculations such bulk simulations can be 
easily performed within the bulk regime, thus being rather simple and reliable.
In addition, by observing the topological phase transition in the bulk provides 
a wide range of new information regarding the scenario occurring on the borderline 
between non-trivial and trivial systems, where the mixing and interdiffusion 
often takes place in the quantum well structure formed by these two materials~\cite{OZO+01,OZP+02}.\\
\indent
In the following example we will consider the sequence of random alloys
between the prototypical non-trivial gapless semiconductor HgTe and the trivial
ZnS insulator, which are both non-magnetic binaries of the zinc-blende type.
Since Zn is isovalent to Hg and S to Te, the intermediate alloys
(Hg$_{1-x}$Zn$_x$)(Te$_{1-x}$S$_x$) must be non-magnetic semiconductors,
thus keeping the time-reversal symmetry of the total wave function. 
Although here we are investigating the aforementioned alloy series, we have to 
point out that all features of the topological phase transition shown in the
current work can be transferred with no restriction to other systems,
like for e.\,g.\@ the rich family of the ternary Heusler semiconductors where
many topologically non-trivial compounds where already
identified~\cite{CQK+10}.
%------------------------------------------------------------------------------
\section{Computational methodology}
\label{sec:comput}
%------------------------------------------------------------------------------
Until now the CPA (and its extensions) remains the only widely applicable
alloy technique which incorporates the effects of the energy-dependent shift and
lifetime broadening. Both are essential features of the electron localization caused by 
chemical disorder, which are not accessible via other theories, as e.\,g.\@ 
the VCA (virtual crystal approximation)~\cite{Nor31} or supercell calculations.
Although the CPA method has lots of advantages, it has a typical shortcoming,
namely in its original ``single-site'' formulation the effect of the local
environment is absent.
This bottleneck of CPA can be remediated, however, by employing its non-local
extensions~\cite{MSC83,VB00,RSG03}.
Regardless, we would like to emphasize that for the case of isovalent 
substitution the single-site CPA remains a quite good approximation even in the
diluted limit, since the isovalent atoms intermix without additional
environmental preference as long as the zinc-blende type crystal structure
is preserved.\\
\indent
All electronic structure and spin-Hall transport coefficient calculations
presented below were carried out using the fully-relativistic
Korringa-Kohn-Rostoker (KKR) Green's function method implemented within the 
SPR-KKR package~\cite{EKM11} employing the density functional theory framework.
The exchange and correlation was treated using the Vosko-Wilk-Nusair
form of the local density approximation (LDA)~\cite{VWN80}.
LDA is known for its typical underestimation of the band gaps of semiconductors.
However, in case of our study this issue is not critical, since the observed 
trends are qualitatively well described even with LDA.
The lattice constants for the alloys were derived by linear interpolation 
between the experimental lattice parameters of the pure HgTe and ZnS compounds
as a function of the concentration of the constituents.
Due to the multiple-scattering construction of the Green's function~\cite{FS80}, 
the method provides a suitable base for mean-field approaches like CPA. 
Since the electronic structure of disordered materials in general does not
show well defined Bloch eigenstates, their electronic structure is described
by the Bloch-spectral function (BSF) defined by the imaginary part
of the alloy Green's function that is diagonal in momentum space~\cite{FS80}.
%------------------------------------------------------------------------------
\section{Results and discussion}
\label{sec:result}
\begin{figure}
  \centering
  \includegraphics[width=0.85\linewidth]{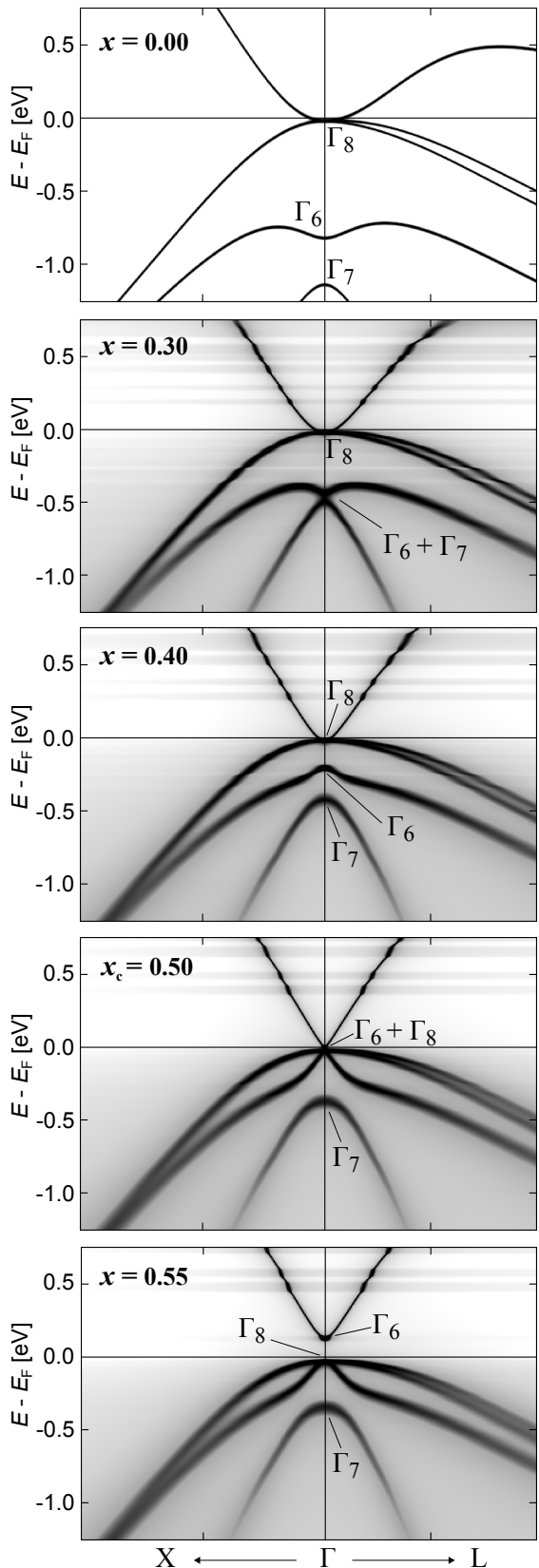}
  \caption{Evolution of the Bloch-spectral function (along ${X\!-\!\Gamma\!-\!L}$) 
   in the alloy series (Hg$_{1-x}$Zn$_x$)(Te$_{1-x}$S$_x$). 
   The critical composition ${x_{\rm c}=0.5}$ corresponds to the topological phase 
   transition marked by a Dirac cone.}
  \label{fig:bns}
\end{figure}

\begin{figure}[t]
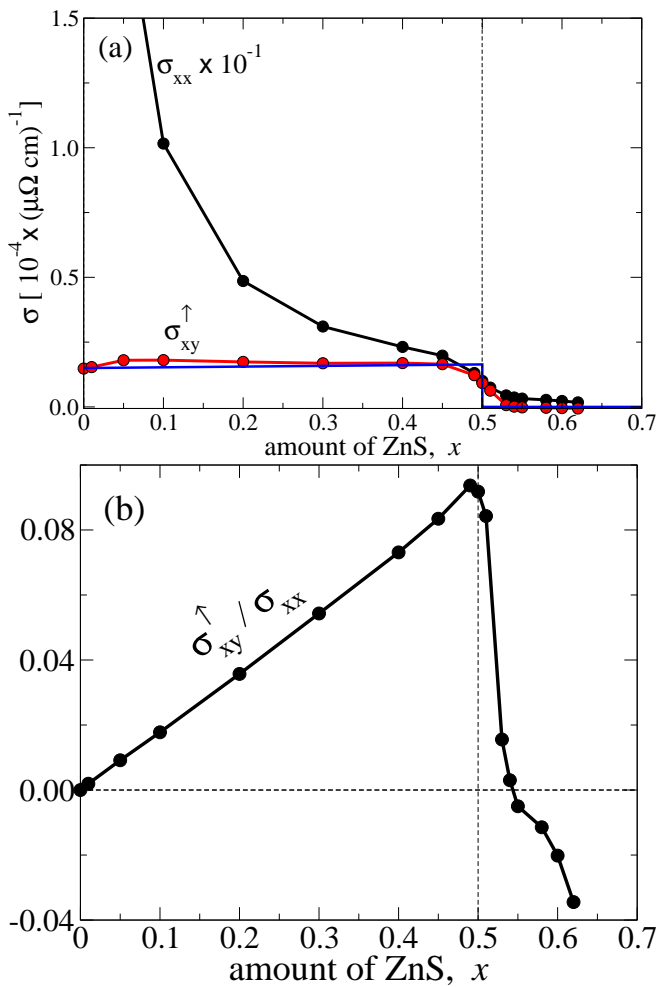

  \centering
  \includegraphics[clip,width=1.0\linewidth]{figure2a.eps}\\
  \includegraphics[clip,width=1.0\linewidth]{figure2b.eps}
  \caption{(color online) (a) Diagonal ($\sigma_{\rm xx}$, black) and spin-projected
    off-diagonal ($\sigma^{\uparrow}_{\rm xy}$, red) components of
    the conductivity tensor calculated as a function of $x$ 
    in the disordered (Hg$_{1-x}$Zn$_x$)(Te$_{1-x}$S$_x$) alloy series.
    These results are compared to the quantum of the 3D spin-Hall
    conductivity marked by the blue line.
    (b) The spin-Hall ratio $\sigma^{\uparrow}_{\rm xy}$/$\sigma_{\rm xx}$ 
    plotted as a function of composition $x$.}
  \label{fig:sig}
\end{figure}

%------------------------------------------------------------------------------
In order to study the topological phase transition in bulk 3D materials,
we have performed band structure calculations and we have investigated the
spin-resolved transport properties of several quaternary alloys in the series 
of (Hg$_{1-x}$Zn$_x$)(Te$_{1-x}$S$_x$).
First, we present the analysis of the BSF computed for the most relevant alloy
compositions, which are shown in Fig.\,\ref{fig:bns}.
%%%%%%%%%%%%%%%%%%%%%%%%%%%%%%%%%%%%%%%%%%%%%%%%%%%%%%%
%%%%%%%%%%%%%%%%%%%%%%%%%%%%%%%%%%%%%%%%%%%%%%%%%%%%%%%
Since the band structure of pure HgTe (${x=0}$) was established long
ago~\cite{BP76}, here we just briefly mention the most characteristic features
in the vicinity of the $\Gamma$ point:
the Te $5p$-band is split due to spin-orbit coupling (SOC) into a $\Gamma_8$
state sitting right at $E_{\rm F}$ and a $\Gamma_7$ state, which is located about 
1.2~eV below $\Gamma_8$. 
The Hg $6s$-band corresponding to $\Gamma_6$ is placed by the crystal field (CF) 
between the two aforementioned states, at about -0.8~eV from $E_{\rm F}$.
This negative value, often named ``band inversion'', drives HgTe into a non-trivial
class, resulting from the combined effect of the strong SOC and small CF.
For ${x>0}$ the symmetry analysis cannot be applied directly, since
there are no pure Bloch eigenstates, i.\,e.\@  in the disordered regime the
delta-like poles of the BSF turn into overlapping Lorentzians spread over the 
whole $(\vec{k},E)$-space. 
However, it is still interesting to trace the evolution of their maxima in analogy
to the dispersion relation $E(\vec k)$ of the pure material.
By calculations we will show that the SOC and CF can be tuned as a function
of composition~\cite{CQK+10}.
Upon gradual substitution of Hg with Zn, and Te with S we effectively decrease
the SOC and increase the CF. Hence, both $\Gamma_7$- and $\Gamma_6$-like states rise in energy 
(see Fig.\,\ref{fig:bns}). In the ZnS-poor regime the decrease of SOC acts very efficiently on the BSF,
and at about ${x=0.3}$ the $\Gamma_7$-like state touches the $\Gamma_6$-like
state at roughly -0.5~eV below $E_{\rm F}$. As far as there is no change of occupancies this does not affect the topological 
class of the system. By further substitution the $\Gamma_6$ state continues to rise towards 
$E_{\rm F}$, being strongly influenced by the CF, whereas the effective decrease
of SOC does not effect the $\Gamma_7$-like state considerably.
As shown by the snapshot at the intermediate concentration ${x=0.4}$,
the band alignment in this regime is equivalent to the original pure HgTe case.\\
\indent
At the ``critical'' concentration of the random substitution ${x_{\rm c}=0.5}$ 
the system undergoes a topological phase transition manifested by the Dirac cone
formed by a mixture of $\Gamma_8$- and $\Gamma_6$-like states.
Obviously the edge states with linear dispersion at the surface of a topological
insulator and the states forming the Dirac cone in the disordered bulk manifest
the same transition mechanism.
The only qualitative difference is that due to the bulk translational symmetry, 
we observe two replica of Dirac cones with opposite spins superimposed, whereas 
due to the break of space-reversal symmetry at the surface only a single cone 
remains, which exposes the adiabatic spin-current.\\
\indent
In the ZnS-rich regime (${x>0.5}$) which is topologically trivial
due to the emptying of $\Gamma_6$-symmetric state, the evolution of the energy
levels are basically unaffected by~$x$: $\Gamma_7$ is pinned to $\Gamma_8$
at $E_{\rm F}$ due to vanishing SOC, and the unoccupied $\Gamma_6$ state
continues to move up increasing the band gap width. It is easy to see that
in this way we will arrive to the band structure of pure ZnS, which is rather
well known~\cite{RL66,EMT67}.
Within this regime the electronic structure is close to the topologically 
``intermediate'' pure compounds similar to CdTe~\cite{BP76} or
CdSe~\cite{MNHE98}, which are close to the borderline of trivial and 
topological insulators.\\
\indent
The energy distribution of the lifetime broadening can be understood 
by noticing that it scales with the band curvature 
\mbox{$\sim\!\partial^2 E/\partial k^2$}, i.\,e.\@ with the effective mass.
Indeed, the corresponding shades represent a weighted superposition of 
parabolic-like heavy bands exhibited by both pure HgTe and ZnS.
In contrast to this, the appearance of Bloch-like states with linear dispersion,
--i.\,e.\@ the massless states-- are essentially a new ordering feature arising
from the quantum interference. This situation is very evident for the critical
composition $x_{\rm c}$=0.5 which corresponds to the appearance of the
conical dispersion and simultaneously most strongly influenced by disorder.\\
\indent
To verify that this is indeed a topological phase transition, in the following
we will analyze the tensor of residual conductivity 
calculated by means of the general Kubo and Kubo-St\v{r}eda
formalism~\cite{But85,LGK+11}. In order to access the spin-resolved
components $\sigma_{\rm xy}^{\uparrow(\downarrow)}$ of the transversal
conductivity, where the spin polarization $\uparrow(\downarrow)$ refers
to the $z$-axis, we employed a relativistic scheme suggested 
recently~\cite{VGW07,LGK+11}. As it follows from Fig.\,\ref{fig:sig}~(a), the diagonal component $\sigma_{\rm xx}$
%%%%%%%%%%%%%%%%%%%%%%%%%%%%%%%%%%%%%%%%%%%%%%%%%%%%%%%
%%%%%%%%%%%%%%%%%%%%%%%%%%%%%%%%%%%%%%%%%%%%%%%%%%%%%%%
scales almost as ${\sim\!x^{-1}}$ from infinity (in the pure
limit HgTe is a metal in our calculations) down to the values, which within a 
strong disorder regime become already comparable with the spin-Hall component
$\sigma^{\uparrow}_{\rm xy}$.
On the ZnS-rich side (${x>0.5}$) both components quickly drop down almost to zero
as a function of $x$, since the band gap starts to open. 
Due to the delocalization which takes place both in the $\vec{k}$- as well as 
in the energy space, this drop is not abrupt. 
Moreover, we notice that $\sigma^{\uparrow}_{\rm xy}$ does change its sign 
before it vanishes, which can be seen from the dependency of spin-Hall ratios 
(shown in Fig.\,\ref{fig:sig}~(b)). 
The sign change is connected to the reversal of the Hall chirality 
after the cone center crosses the Fermi energy $E_{\rm F}$.\\
\indent
In contrast to the diagonal component, the spin-Hall conductivity stays almost
independent from the composition until the topological phase transition takes place,
which indicates the quantized origin of $\sigma^{\uparrow}_{\rm xy}$.
Indeed, since $\sigma^{\uparrow}_{\rm xy}$ is calculated within the Kubo-formalism,
together with the intrinsic quantized contribution 
(connected with the Berry curvature), it contains also extrinsic
contributions (attributed to the side-jump and skew scattering~\cite{NSO+10,LGK+11}
mechanisms). However, for the present alloy series which behave almost as 
gapless semiconductors within the whole topologically non-trivial
regime, the extrinsic contributions must be small, since the DOS at the
Fermi energy, which supplies the electrons for the corresponding scattering is very
small as well.   As it follows from Fig.\,\ref{fig:sig}~(a), within topologically
non-trivial regime the value of \mbox{$\sigma^{\uparrow}_{\rm xy}$} 
 reasonably agrees with the quantum of the spin-Hall conductivity in 
3D~\cite{MK90,KHW93} (marked by a blue line in Fig.\,\ref{fig:sig}~(a)).
The latter is calculated as $\frac{1}{2}\frac{e^2}{2\pi h}K$,
where ${K=\pi/a}$ is the size of the magnetic Brillouin zone in $z$ direction. 
The factor $\nicefrac{1}{2}$ accounts for the single spin-channel.
Since the lattice constant $a$ decreases as a function of $x$, the
quantum of the conductivity slightly increases with  rising Hg concentration 
until the topological phase transition. 

%------------------------------------------------------------------------------
\section{Conclusions}
\label{sec:conc}
%------------------------------------------------------------------------------
In this paper we have shown that the topological phase transition induced via 
disorder can be efficiently studied in bulk 3D materials, without the
need to investigate large systems, circumventing computationally quite heavy 
surface calculations.
The main conclusion of our study is based on the composition dependence of the
spin-Hall ratio $\sigma_{\rm xy}^{\uparrow}/\sigma_{\rm xx}$ shown in 
Fig.\,\ref{fig:sig}~(b): the ratio nearly  scales linearly up with increasing 
disorder in the system. This behavior can be exploited as an efficient tool for
adjusting the adiabatic transport characteristics. 
The upper limit of such an adjustment can obviously
be achieved in the case of full Anderson localization, i.\,e.\@ 
when ${\sigma_{\rm xx}=0}$. The concept of the Anderson type topological
insulator (TAI) joins two very fundamental fields and it is not surprising that 
it immediately became a subject of intensive theoretical research. 
The possibility of the TAI state was indicated by C.~Beenakker: 
{\em Anderson insulator exists due to disorder, whereas topological insulator 
exists in spite of disorder.}
On a model level, provided that the Anderson localization exists, the possibility 
for the TAI state  has already been justified~\cite{LCJS09,GWA+09,JWSX09}.
However, even though the Anderson model of localization is rather
simple~\cite{And58,Tho74}, the question whether the full localization can
be adequately described in 3D solids within CPA-like theories remains the subject 
of vivid debates~\cite{EC72,MP72,Bis73,KK86,KSW93,SE99,MC05}.

\begin{acknowledgments}
\noindent
Financial support by the DFG projects FOR~1464 ``$\text{ASPIMATT}$'' and SFB~689 ``Spinph\"anomene in reduzierten
Dimensionen'' is gratefully acknowledged.
\end{acknowledgments}

%\bibliography{database}

\end{document}